\documentclass[twocolumn,tighten]{aastex7}
\usepackage{color}
\newcommand{\red}{\textcolor{black}}
\usepackage{multirow}
\shortauthors{Sano, Fukui \& Fujimori et al. (2026)}
\usepackage{comment}

\begin{document}
\title{Discovery of Molecular and Atomic Gas associated with HESS~J1646$-$458 (Westerlund~1):\\Spatial TeV Gamma-Ray and Interstellar Proton Correspondence}

\shorttitle{Molecular and Atomic Gas associated with HESS~J1646$-$458 (Westerlund~1)}

\author[0000-0003-2062-5692]{H. Sano}
\affiliation{Faculty of Engineering, Gifu University, 1-1 Yanagido, Gifu 501-1193, Japan}
\affiliation{Department of Intelligence Science and Engineering, Graduate School of Natural Science and Technology, Gifu University, 1-1 Yanagido, Gifu, 501-1193 Japan}
\affiliation{Center for Space Research and Utilization Promotion (c-SRUP), Gifu University, 1-1 Yanagido, Gifu 501-1193, Japan}
\email[show]{sano.hidetoshi.w4@f.gifu-u.ac.jp}  

\author[0000-0002-8966-9856]{Y. Fukui}
\affiliation{Faculty of Engineering, Gifu University, 1-1 Yanagido, Gifu 501-1193, Japan}
\affiliation{Department of Physics, Nagoya University, Furo-cho, Chikusa-ku, Nagoya 464-8601, Japan}
\email[show]{\\\hspace*{0.6cm}fukui@a.phys.nagoya-u.ac.jp}  

\author[0009-0008-1453-5459]{S. Fujimori}
\affiliation{Faculty of Engineering, Gifu University, 1-1 Yanagido, Gifu 501-1193, Japan}
\email[show]{\\\hspace*{0.6cm}fujimori.shota.t5@s.gifu-u.ac.jp} 

\author[0000-0002-9552-3570]{T. Murase}
\affiliation{Faculty of Engineering, Gifu University, 1-1 Yanagido, Gifu 501-1193, Japan}
\affiliation{Center for Space Research and Utilization Promotion (c-SRUP), Gifu University, 1-1 Yanagido, Gifu 501-1193, Japan}
\email{murase.takeru.b0@f.gifu-u.ac.jp}

\author[0000-0001-5609-7372]{R. Z. E. Alsaberi}
\affiliation{Faculty of Engineering, Gifu University, 1-1 Yanagido, Gifu 501-1193, Japan}
\affiliation{Western Sydney University, Locked Bag 1797, Penrith South DC, NSW 1797, Australia}
\email{rami.zainal.ezzat.alsaberi.j1@f.gifu-u.ac.jp}  

\author[0000-0002-4990-9288]{M. D. Filipovi{\'c}}
\affiliation{Western Sydney University, Locked Bag 1797, Penrith South DC, NSW 1797, Australia}
\email{m.filipovic@westernsydney.edu.au}

\author[0000-0002-9516-1581]{G. Rowell}
\affiliation{School of Physical Sciences, The University of Adelaide, North Terrace, Adelaide, SA 5005, Australia}
\email{gavin.rowell@adelaide.edu.au}

\author[0000-0001-5069-5988]{M. Aruga}
\affiliation{Faculty of Engineering, Gifu University, 1-1 Yanagido, Gifu 501-1193, Japan}
\affiliation{Department of Physics, Nagoya University, Furo-cho, Chikusa-ku, Nagoya 464-8601, Japan}
\email{aruga.maki.d4@s.mail.nagoya-u.ac.jp}

\author[0009-0004-3558-3477]{Y. Asano}
\affiliation{Faculty of Engineering, Gifu University, 1-1 Yanagido, Gifu 501-1193, Japan}
\email{asano.yuuya.m9@s.gifu-u.ac.jp}

\author[0000-0002-0160-8865]{R. G. Bhuvana}
\affiliation{Faculty of Engineering, Gifu University, 1-1 Yanagido, Gifu 501-1193, Japan}
\affiliation{Institute for Advanced Study, Gifu University, 1-1 Yanagido, Gifu 501-1193, Japan}
\email{bhuvana.g.r.k4@f.gifu-u.ac.jp}

\author[0009-0002-0025-1646]{F. Demachi}
\affiliation{Department of Physics, Nagoya University, Furo-cho, Chikusa-ku, Nagoya 464-8601, Japan}
\email{f.demachi@a.phys.nagoya-u.ac.jp}  

\author[0000-0001-9687-8237]{S. Einecke}
\affiliation{School of Physical Sciences, The University of Adelaide, North Terrace, Adelaide, SA 5005, Australia}
\email{sabrina.einecke@adelaide.edu.au}

\author[0009-0005-4458-2908]{N. Fukaya}
\affiliation{Department of Physics, Nagoya University, Furo-cho, Chikusa-ku, Nagoya 464-8601, Japan}
\email{n.fukaya@a.phys.nagoya-u.ac.jp}  

\author[]{R. Hamada}
\affiliation{Faculty of Engineering, Gifu University, 1-1 Yanagido, Gifu 501-1193, Japan}
\affiliation{Department of Physics, Nagoya University, Furo-cho, Chikusa-ku, Nagoya 464-8601, Japan}
\email{hamada.riku.p8@s.mail.nagoya-u.ac.jp}

\author[0009-0003-8705-5695]{H. Inoue}
\affiliation{Faculty of Engineering, Gifu University, 1-1 Yanagido, Gifu 501-1193, Japan}
\email{inoue.haruto.s4@s.gifu-u.ac.jp}

\author[]{T. Kamazaki}
\affiliation{National Astronomical Observatory of Japan (NAOJ), National Institutes of Natural Sciences (NINS), 2-21-1, Osawa, Mitaka, Tokyo 181-8588, Japan}
\email{kamazaki.takeshi@nao.ac.jp}  

\author[0000-0001-6109-8548]{S. Lazarevi\'c}
\affiliation{Western Sydney University, Locked Bag 1797, Penrith South DC, NSW 1797, Australia}
\affiliation{CSIRO Space and Astronomy, Australia Telescope National Facility, PO Box 76, Epping, NSW 1710, Australia}
\affiliation{Astronomical Observatory, Volgina 7, 11060 Belgrade, Serbia}
\email{s.lazarevic@westernsydney.edu.au}

\author[0000-0001-9778-6692]{T. Minamidani}
\affiliation{National Astronomical Observatory of Japan (NAOJ), National Institutes of Natural Sciences (NINS), 2-21-1, Osawa, Mitaka, Tokyo 181-8588, Japan}
\affiliation{Graduate Institute for Advanced Studies, SOKENDAI, 2-21-1, Osawa, Mitaka, Tokyo 181-8588, Japan}
\email{tetsuhiro.minamidani@nao.ac.jp}  

\author[0009-0009-7061-0553]{Z. J. Smeaton}
\affiliation{Western Sydney University, Locked Bag 1797, Penrith South DC, NSW 1797, Australia}
\email{19594271@student.westernsydney.edu.au}

\author{H. Sudou}
\affiliation{National Institute of Technology, Sendai College 48 Nodayama, Medeshima-Shiote, Natori, Miyagi 981-1239, Japan}
\email{sudo-hiroshi@sendai-nct.ac.jp}

\author[0000-0002-1411-5410]{K. Tachihara}
\affiliation{Department of Physics, Nagoya University, Furo-cho, Chikusa-ku, Nagoya 464-8601, Japan}
\email{k.tachihara@a.phys.nagoya-u.ac.jp}  

\author{H. Takaba}
\affiliation{Faculty of Engineering, Gifu University, 1-1 Yanagido, Gifu 501-1193, Japan}
\affiliation{National Astronomical Observatory of Japan (NAOJ), National Institutes of Natural Sciences (NINS), 2-21-1, Osawa, Mitaka, Tokyo 181-8588, Japan}
\email{hiroshi.takaba@nao.ac.jp}

\author[0000-0002-2794-4840]{K. Tsuge}
\affiliation{Faculty of Engineering, Gifu University, 1-1 Yanagido, Gifu 501-1193, Japan}
\affiliation{Institute for Advanced Study, Gifu University, 1-1 Yanagido, Gifu 501-1193, Japan}
\affiliation{National Astronomical Observatory of Japan (NAOJ), National Institutes of Natural Sciences (NINS), 2-21-1, Osawa, Mitaka, Tokyo 181-8588, Japan}
\email{tsuge.kisetsu.i2@f.gifu-u.ac.jp}  

\author[0000-0002-1865-4729]{R. I. Yamada}
\affiliation{Faculty of Engineering, Gifu University, 1-1 Yanagido, Gifu 501-1193, Japan}
\affiliation{National Astronomical Observatory of Japan (NAOJ), National Institutes of Natural Sciences (NINS), 2-21-1, Osawa, Mitaka, Tokyo 181-8588, Japan}
\email{yamada.rin.x6@f.gifu-u.ac.jp}

\correspondingauthor{H. Sano, Y. Fukui, \& S. Fujimori}

\begin{abstract}
We \red{report} CO and H{\sc i} studies of molecular and atomic gas toward the TeV gamma-ray source HESS~J1646$-$458, widely considered to be associated with the young massive cluster Westerlund~1 (Wd1). We found \red{that} molecular clouds at $V_\mathrm{LSR} \sim$$-32$~km~s$^{-1}$ coincide with arc-like structures seen at 8~$\mu$m, likely illuminated by strong FUV radiation from Wd1. $^{12}$CO($J$~=~3--2) emission at the same velocity reveals a cavity-like structure with an expansion velocity of $\sim$$5$~km~s$^{-1}$ toward the central region of Wd1, suggesting a recently formed wind-blown bubble driven by the cluster. We also identify a complementary spatial distribution between the $V_\mathrm{LSR} \sim$$-55$ and $\sim$$-32$~km~s$^{-1}$ clouds, connected by an intermediate-velocity component at $V_\mathrm{LSR} \sim$$-44$~km~s$^{-1}$. These characteristics are consistent with signatures of triggered star formation through a cloud-cloud collision and imply that both clouds are physically associated with Wd1. On larger scales, the total interstellar proton column density at $V_\mathrm{LSR}$ $\sim$$-36$--$-23$~km~s$^{-1}$ shows a moderate spatial correspondence with the TeV gamma-ray shell. Together with this correlation, a substantial gas mass of $\sim$$1.6 \times 10^6~M_\odot$, and the absence of bright synchrotron X-rays, the TeV gamma-ray emission surrounding Wd1 \red{is consistent with} the hadronic origin. The present finding allows us to calculate the total energy of accelerated cosmic-ray protons to be $\sim$$6 \times 10^{49}$~erg.
\end{abstract}


\section{Introduction}
It is a long-standing issue how cosmic rays (CRs), mainly consisting of relativistic protons, are accelerated in interstellar space. Supernova remnants (SNRs) are believed to be primary accelerators of Galactic CRs up to $\sim$3~PeV (knee energy) via diffusive shock acceleration \citep[e.g.,][]{1978MNRAS.182..147B,1978ApJ...221L..29B}. The detection of GeV gamma-rays with the pion-decay bump (known as ``hadronic gamma-rays'') confirmed that at least 10~GeV CR protons are accelerated in the middle-aged SNRs \citep[e.g.,][]{2011ApJ...742L..30G,2013Sci...339..807A}. For the high-energy end, a good spatial correspondence between the TeV gamma-rays and interstellar protons in young SNRs provides compelling evidence for CR proton acceleration up to $\sim$100~TeV because the hadronic gamma-ray flux is proportional to the interstellar proton density \citep[e.g.,][]{2012ApJ...746...82F,2017ApJ...850...71F}. Despite these extensive efforts, no SNR has been reported that accelerates CR protons close to the knee energy.

Young massive clusters bright in gamma-rays have received much attention because of their potential for accelerating CRs close to the knee energy since the pioneering study by \cite{1979ApJ...231...95M}. The CR proton acceleration could operate in the vicinity of massive stars via strong stellar winds \citep[e.g.,][]{1980ApJ...237..236C,1983SSRv...36..173C} and/or inside a superbubble \citep[][and references therein]{2014A&ARv..22...77B}. Theoretically, the multiple shocks have the potential to induce an efficient acceleration of CR protons above 1~PeV \citep[e.g.,][]{2000APh....13..161K}. Several Galactic and Magellanic young massive clusters and superbubbles have been detected in GeV/TeV gamma-rays that could be due to the hadronic process: the decay of neutral pions produced when CR protons interact with interstellar protons \citep[e.g.,][]{2011Sci...334.1103A,2012A&A...537A.114A,2017A&A...600A.107Y,2015Sci...347..406H}. One current challenge is to evaluate the CR proton acceleration and its energy in the gamma-ray bright young massive clusters or superbubbles observationally. 

Westerlund~1 (Wd1), located at ($l$, $b$) $\sim$ (339\fdg55, $-$0\fdg40) or ($\alpha_\mathrm{J2000}$, $\delta_\mathrm{J2000}$) $\sim$ ($16^\mathrm{h}47^\mathrm{m}02\fs4$, $-45\arcdeg51\arcmin07\arcsec$), is the most massive stellar cluster in the Milky Way with an estimated total stellar mass of approximately $5 \times10^4$ to over $10^5$~$M_{\odot}$, containing at least 166 massive stars with an initial mass of $\sim$25--50$M_{\odot}$ \citep[e.g.,][]{2005A&A...434..949C,2020A&A...635A.187C,2006MNRAS.372.1407C,2006ApJ...636L..41M,2008A&A...478..137B,2010A&A...516A..78N,2011MNRAS.412.2469G}. The age and distance of the cluster are still being debated. Most age estimates suggest a young age of $\sim$3--5~Myrs \citep[e.g.,][]{2005A&A...434..949C,2006MNRAS.372.1407C,2008A&A...478..137B}, although the cluster contains an older population with an age of $\sim$10~Myr \citep[][]{2021ApJ...912...16B,2022MNRAS.516.1289N}. Previous studies have widely accepted a distance of $\sim$3.9~kpc to Wd1 \citep[][]{2005A&A...434..949C,2006MNRAS.372.1407C,2008A&A...478..137B}. Several proper motion measurements based on Gaia data have yielded broadly consistent results, suggesting that the cluster is located in the Norma arm \citep[][]{2019MNRAS.486L..10D,2020MNRAS.495.1209R,2021ApJ...912...16B,2022A&A...664A.146N,2022MNRAS.516.1289N}. On the other hand, recent proper motion analyses using Gaia DR2 and EDR3, based on Bayesian inference, have suggested a closer distance of $\sim$2.6--2.8~kpc \citep[][]
{2020MNRAS.492.2497A,2021RNAAS...5...14A}, and the issue remains under debate. In this paper, we adopt a distance of 3.9 kpc, which is consistent with the majority of previous studies.

Wd1 has attracted attention as a potential CR accelerator following the detection of diffuse GeV/TeV gamma-ray emission. \cite{2012A&A...537A.114A} discovered extended gamma-ray emission named HESS~J1646$-$458 based on 45~hour observations using the High Energy Stereoscopic System (H.E.S.S.). The angular extent of HESS~J1646$-$458 is $\sim$2$^{\circ}$ centered approximately on the geometric center of Wd1. Using the gamma-ray properties in the 0.45--20~TeV range and multiwavelength data, the authors concluded that Wd1 is likely accelerating CR protons and that the observed bulk emission favors to be explained by a hadronic process. Subsequent follow-up observations using Fermi \citep[$E$: 3--300~GeV,][]{2013MNRAS.434.2289O} and H.E.S.S. \citep[exposure increased to 164 hours,][]{2022A&A...666A.124A} also support this scenario. It is worth noting that the gamma-ray spectrum, which shows no significant cutoff, has the potential to extend up to knee energies \citep[c.f.][]{2019RLSFN..30S.159Y}.

In contrast, the hadronic origin of HESS~J1646$-$458 remains uncertain, as a clear spatial correlation between the gamma-rays and interstellar gas has not been established. \cite{2010ApJ...713L..45L} compared the preliminary H.E.S.S. gamma-ray map with the CfA $^{12}$CO($J$~=~1--0) data at an angular resolution of 8\farcm8. They found an extended ring-like CO feature at $V_\mathrm{LSR}$ $\sim$$-90$~km~s$^{-1}$, which is spatially anti-correlated with the gamma-ray distribution. Subsequently, \cite{2012A&A...537A.114A} compared the gamma-rays, CO, and H{\sc i} at $V_\mathrm{LSR}$ $\sim$$-55$~km~s$^{-1}$, corresponding to the kinematic distance of $\sim$3.9~kpc. However, no spatial correlation was found between the CO/H{\sc i} distributions and the gamma-ray peaks. \cite{2022A&A...666A.124A} similarly found no interstellar gas clearly associated with HESS~J1646$-$458, based on their reanalysis of archival CO and H{\sc i} data covering $|b| < 1^{\circ}$ for three velocity components at $V_\mathrm{LSR}$ $\sim$$-55$, $-44$, and $-32$~km~s$^{-1}$.

Here, we report compelling evidence for molecular and atomic gas associated with both HESS~J1646$-$458 and Wd1, based on new NANTEN2 and the Atacama Submillimeter Telescope Experiment (ASTE) observations in addition to the reanalysis of archival CO/H{\sc i} datasets. The moderate spatial correspondence between the gamma-rays and interstellar gas is expected to provide new observational constraints on the origin of HESS~J1646$-$458 and, more broadly, on our understanding of CR accelerators in the Galaxy. Section~\ref{observations} describes observations and data reductions of CO, H{\sc i}, gamma-rays, and other wavelength datasets. Section~\ref{results} comprises four subsections: section~\ref{multiwavelength} presents an overview of multiwavelength datasets, section~\ref{co_hi} describes large-scale distributions of CO and H{\sc i}, section~\ref{hidip} shows H{\sc i} intensity depression toward the southern gamma-ray peak, and section~\ref{aste} presents $^{12}$CO($J$~=~3--2) results toward the central region of Wd1. Discussion and conclusions are given in Sections~\ref{discussion} and \ref{conclusions}, respectively.

\section{Observations and Data Reductions}
\label{observations}
\subsection{CO}
\subsubsection{NANTEN2 $^{12}$CO($J$~=~1--0)}
Observations of $^{12}$CO($J$~=~1--0) at 115~GHz were performed from 2012 May to June using the NANTEN2 4~m millimeter radio telescope of Nagoya University located at Pampa La Bola (altitude of 4,865~m), in northern Chile. We carried out Nyquist sampled on-the-fly mapping observations, which covered an area of $5^{\circ} \times 2^{\circ}$ with a range of $337^{\circ} < l <  342^{\circ}$ and $|b| < 1^{\circ}$. The front-end was a 4~K Nb superconductor-insulator-superconductor (SIS) mixer receiver. The system temperature, including the atmosphere, was $\sim$180~K in the double sideband. The back-end was a digital Fourier transform spectrometer with 1~GHz bandwidth dividing 16,384~channels, corresponding to a velocity coverage of $\sim$2,600~km~s$^{-1}$ and a velocity resolution of $\sim$0.16~km~s$^{-1}$~ch$^{-1}$. The final beam size of reduced cube data was $\sim$$180''$ after convolving with a Gaussian kernel. The pointing offset was better than $\sim$$15''$ by checking every three hours. The absolute intensity was evaluated by observing IRAS~16293$-$2422 at ($\alpha_\mathrm{J2000}$, $\delta_\mathrm{J2000}$) = ($16^{\mathrm{h}}32^{\mathrm{m}}23.3^{\mathrm{s}}$, $-24\arcdeg28\arcmin39\farcs2$) \citep{2006AJ....131.2921R}.

To cover an area of $|b| > 1^{\circ}$, we combined archival $^{12}$CO($J$~=~1--0) datasets obtained using the NANTEN 4~m millimeter radio telescope \citep{2004ASPC..317...59M}. The beam size was smoothed to match the FWHM of the NANTEN2 data. The typical noise fluctuations of the final data cube were $\sim$0.33--0.52~K for $|b| < 1^{\circ}$ and $\sim$0.16~K for $|b| > 1^{\circ}$ at a velocity resolution of 1~km~s$^{-1}$.

\subsubsection{ASTE $^{12}$CO($J$~=~3--2)}
In addition, we carried out $^{12}$CO($J$~=3--2) observations at 345~GHz toward the central region of Wd1 using the ASTE 10-m telescope \citep{2004SPIE.5489..763E} to identify which velocity components of the molecular clouds are physically associated with the cluster. The observations were conducted on 2025 July 29, covering a $9' \times 9'$ region including Wd1 with Nyquist-sampled on-the-fly mapping. The front-end was a 2SB SIS mixer-receiver named ``DASH 345.'' The typical system temperature was $\sim$140--160~K in the single sideband. The back-end was an XF-type spectrometer with a velocity resolution of $\sim$0.065~km~s$^{-1}$ per channel and a velocity coverage of $\sim$2141~km~s$^{-1}$ at 350~GHz. The pointing accuracy was checked every 4~hours, and the measurements were kept within an error of $3''$. The absolute intensity calibration was applied by observing RCW~120 centered at ($\alpha_\mathrm{J2000}$, $\delta_\mathrm{J2000}$) $\sim$ ($17^\mathrm{h}12^\mathrm{m}44$\farcs$0$, $-38\degr31\arcmin01\farcs94$) \citep[][]{2022A&A...659A..36K}. We then obtained an average main-beam efficiency of $\sim$0.66. The final beam size and grid size of the data were $\sim$$22''$ and $8''$, respectively. The typical noise fluctuations are $\sim$0.08~K at a velocity resolution of 1~km~s$^{-1}$.

\subsection{H{\sc i}}
We also used the H{\sc i} Southern Galactic Plane Survey H{\sc i} data taken with a combination of the Australia Telescope Compact Array (ATCA) and the Parkes 64~m radio telescope \citep[][]{2005ApJS..158..178M}. The data cube has a beam size of $\sim$$130''$ and a velocity resolution of 0.82~km~s$^{-1}$. The typical noise fluctuations were $\sim$1.3~K at a velocity resolution of 1~km~s$^{-1}$.

\subsection{Gamma-rays}
We used the gamma-ray significance map of HESS~J1646$-$458 ($E  > 0.37$~TeV) published by \cite{2022A&A...666A.124A} to qualitatively compare the gamma-ray distribution with the interstellar gas. The point-spread function of the data has a 68\% containment radius of approximately 0\fdg07. For quantitative analysis, we adopted the best-fit fluxes for 16 regions (each $0\fdg45 \times 0\fdg45$ in size) within HESS~J1646$-$458, as provided in the same paper.

\begin{figure*}[]
\begin{center}
\includegraphics[width=\linewidth,clip]{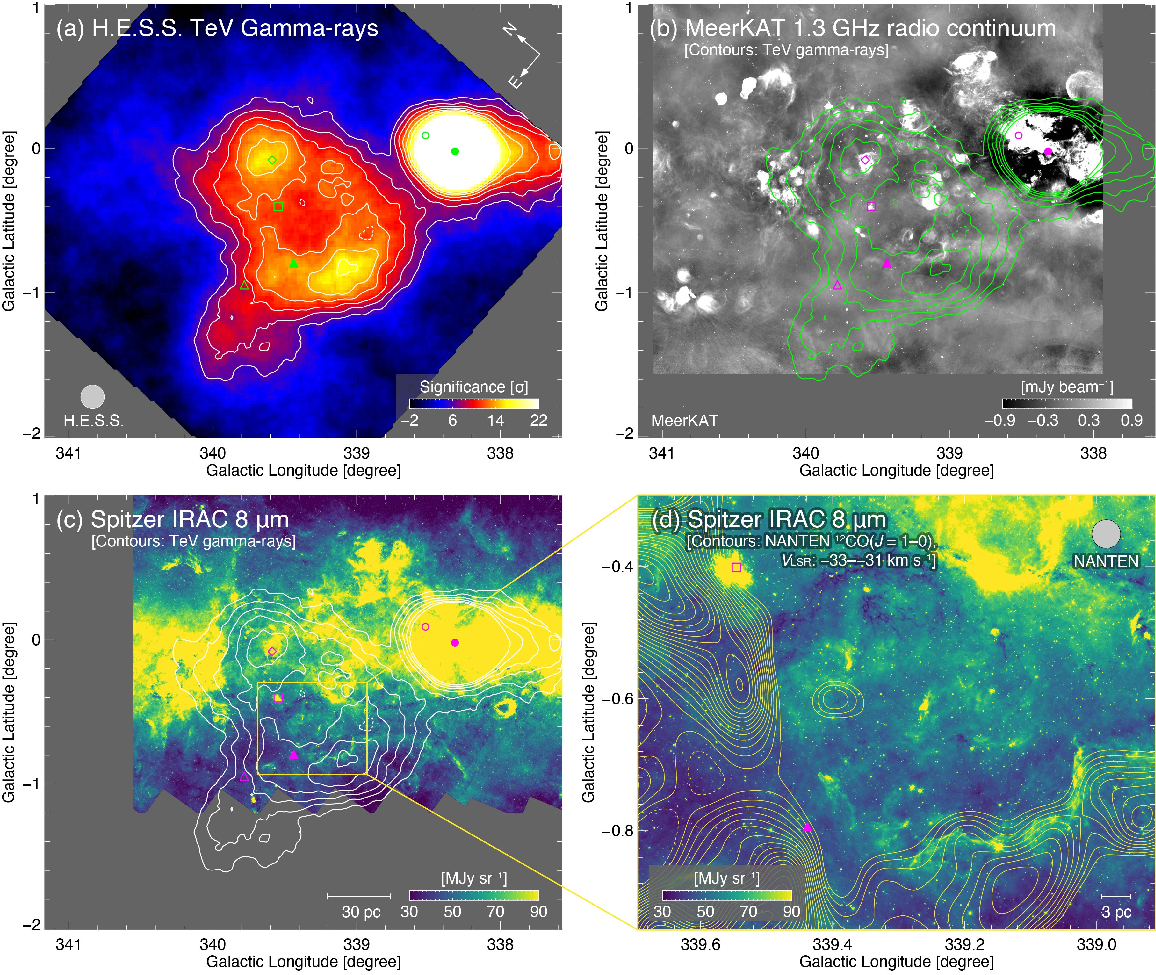}
\caption{(a) TeV gamma-ray significance map of HESS~J1646$-$458 \citep{2022A&A...666A.124A}. The contour levels are 6, 8, 10, 12, 14, and 16$\sigma$. The open square annotates the position of Wd1. The gamma-ray sources HESS~J1641$-$463 and HESS~J1640$-$465 that are not related to HESS~J1646$-$458 are also shown by open and filled circles, respectively. The symbols of open/filled triangles and a diamond represent the positions of PSR~J1650$-$4601, PSR~J1648$-$4611, and 4U~1642$-$45, respectively. (b, c) Maps of (b) MeerKAT radio continuum at 1.3~GHz \citep{2024MNRAS.531..649G} and (c) the Spitzer 8~$\mu$m \citep{2009PASP..121..213C}. Superposed contours are the same as shown in Figure~\ref{fig1}(a). (d) Same as Figure~\ref{fig1}(c), but for an enlarged view near Wd1. Superposed contours indicate the NANTEN $^{12}$CO($J$~=~1--0) intensity \citep{2004ASPC..317...59M}. The integrated velocity range is from $-33$ to $-31$~km~s$^{-1}$. The lowest contour level and contour intervals are 10.5 and 0.5~K~km~s$^{-1}$, respectively.}
\label{fig1}
\end{center}
\end{figure*}%

\begin{figure*}[]
\begin{center}
\includegraphics[width=\linewidth,clip]{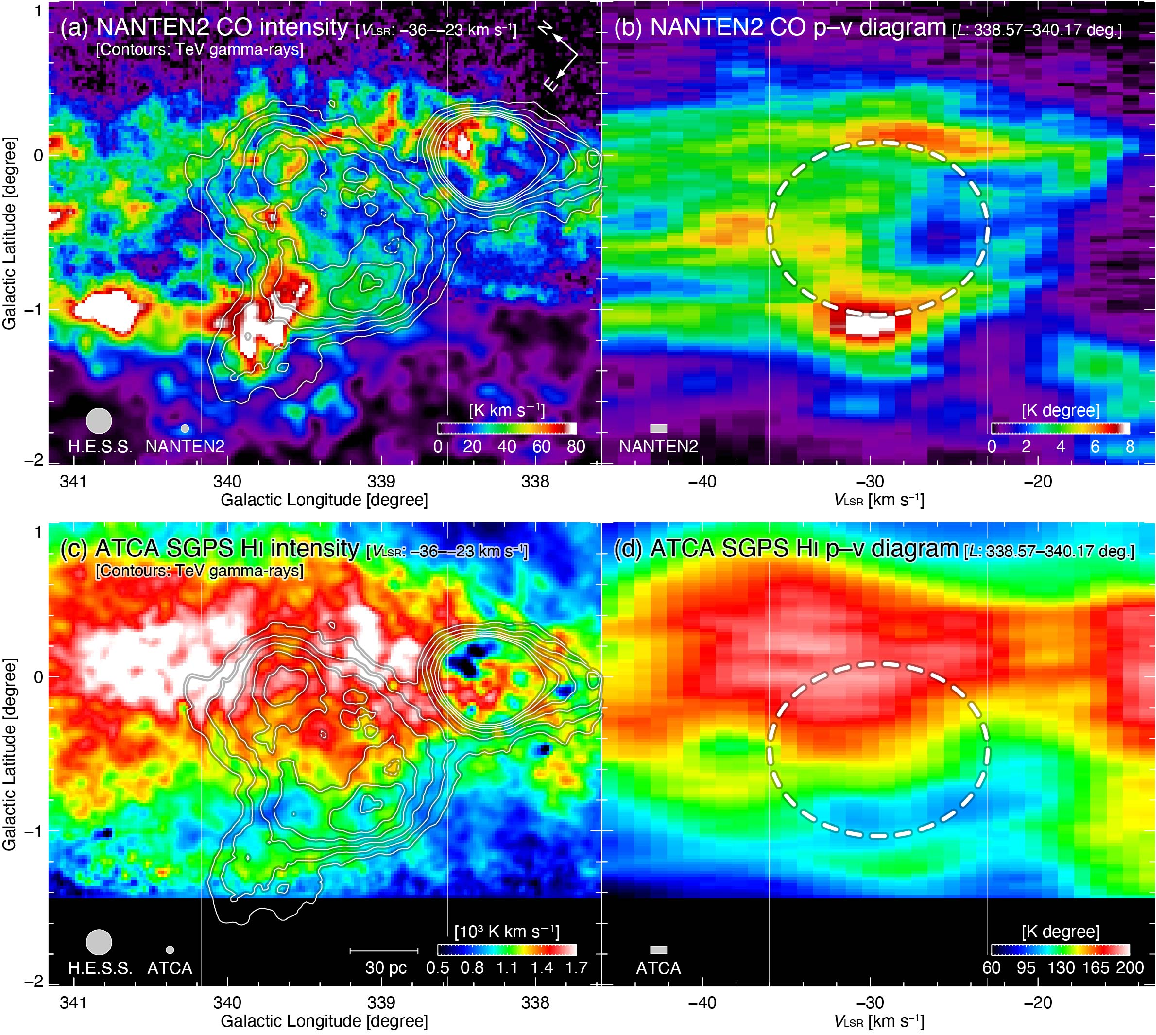}
\caption{Integrated intensity maps and position--velocity ($p$--$v$) diagrams of the NANTEN \& NANTEN2 $^{12}$CO($J$~=~1--0) (a, b) and the ATCA \& Parkes H{\sc i} (c, d). The integration range is from $-36$ to $-23$~km~s$^{-1}$ in velocity for each integrated intensity map and from 338\fdg57 to 340\fdg14 in Galactic Longitude for each $p$--$v$ diagram. The superposed contours in each integrated intensity map are the same as shown in Figure~\ref{fig1}a. The dashed circles in each $p$--$v$ diagram indicate the boundaries of the CO and H{\sc i} expanding shells (see the text).}
\label{fig2}
\end{center}
\end{figure*}%

\begin{figure*}[]
\begin{center}
\includegraphics[width=\linewidth,clip]{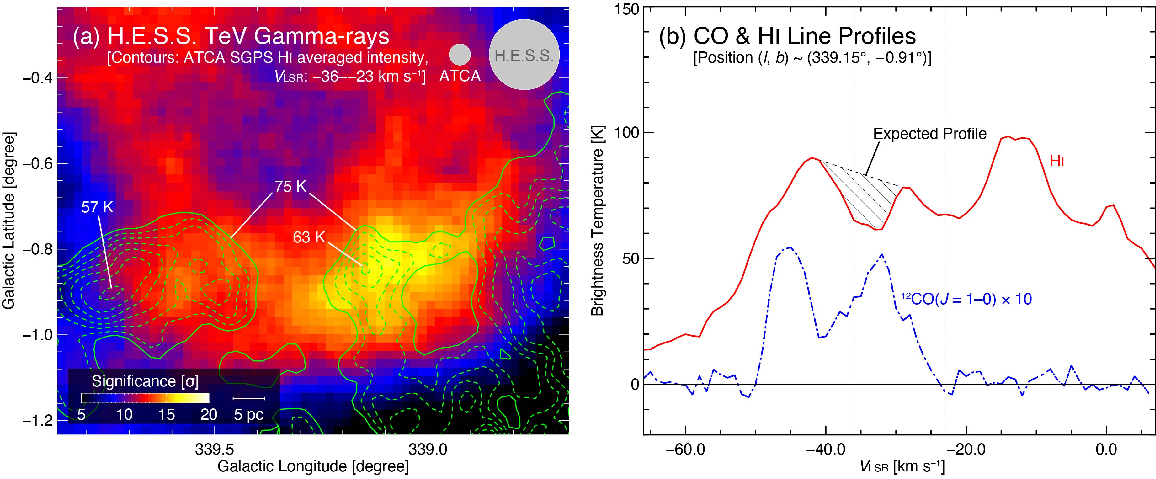}
\caption{(a) Enlarged view of the southern TeV gamma-ray shell. The superposed contours indicate the distribution of averaged H{\sc i} brightness temperature in a velocity range from $-36$ to $-23$~km~s$^{-1}$. (b) Line profiles of CO and H{\sc i} at ($l$, $b$) $\sim$ (339\fdg15, $-$0\fdg91). The CO brightness temperature is scaled by a factor of 10 to match the vertical axis. The vertical dashed lines represent a velocity range from $-36$ to $-23$~km~s$^{-1}$.}
\label{fig3}
\end{center}
\end{figure*}%

\subsection{Other wavelength data}
To investigate the presence of star-forming regions, H{\sc ii} regions, and diffuse radio continuum emission within HESS~J1646$-$458, we used the Spitzer 8~$\mathrm{\mu}$m intensity map from the GLIMPSE survey \citep{2009PASP..121..213C} and the 1.3~GHz MeerKAT radio continuum map from the SARAO MeerKAT Galactic Plane Survey \citep[SMGPS;][]{2024MNRAS.531..649G}. Note that the 8~$\mathrm{\mu}$m image predominantly traces polycyclic aromatic hydrocarbon (PAH) line emission, which is easily excited by FUV radiation from massive stars, while the 1.3~GHz continuum represents both the thermal free-free and non-thermal synchrotron radiation. The angular resolution is $\sim$$2''$ for the Spitzer 8~$\mathrm{\mu}$m and $\sim$$8''$ for the MeerKAT 1.3~GHz continuum.

\section{Results}
\label{results}
\subsection{Multiwavelength Views of HESS~J1646$-$458}
\label{multiwavelength}
Figure~\ref{fig1}(a) shows the distribution of H.E.S.S. TeV gamma-rays toward HESS~J1646$-$458. The most prominent feature in the gamma-ray map is the brightest peak near ($l$, $b$) $\sim$ (338\fdg3, 0\fdg0), corresponding to HESS~J1641$-$463 and HESS~J1640$-$465. However, since this gamma-ray emission is believed to be associated with the unrelated SNRs G338.5$+$0.1 and G338.3$-$0.0 rather than HESS~J1646$-$458, we do not discuss it further in this paper \citep{2014ApJ...794L...1A,2014MNRAS.439.2828A}. HESS~J1646$-$458 itself exhibits a shell-like gamma-ray morphology with peaks at ($l$, $b$) $\sim$ (339\fdg6, 0\fdg0) and (339\fdg1, $-$0\fdg9). Notably, the emission does not peak toward the direction of Wd1, and a blowout-like structure deviating from the shell-like morphology is seen around ($l$, $b$) $\sim$ (339\fdg8, $-$1\fdg2).

Figures~\ref{fig1}(b) and \ref{fig1}(c) show the distributions of MeerKAT 1.3~GHz radio continuum and Spitzer 8~$\mu$m emission. We find no evidence of diffuse radio continuum or infrared emission that traces the full extent of the gamma-ray shell-like structure. On the other hand, we have identified an arc-like feature in the 8~$\mu$m image that corresponds to the southern part of the gamma-ray shell. Figure~\ref{fig1}(d) shows an enlarged view of the 8~$\mu$m image, including the arc-like feature and Wd1. We conducted an unbiased search of the NANTEN CO data over all velocity ranges and found a CO counterpart at $V_\mathrm{LSR}$ $\sim$$-33$--$-31$~km~s$^{-1}$. In the following sections, we mainly focus on the interstellar gas at $V_\mathrm{LSR}$ $\sim$$-32$~km~s$^{-1}$.

\subsection{Distributions of CO and H{\sc i}}
\label{co_hi}
Figures~\ref{fig2}(a) and \ref{fig2}(c) show the velocity-integrated intensity maps of CO and H{\sc i} overlaid with the TeV gamma-ray contours. CO emission is predominantly distributed from the northern to the southeastern parts of the gamma-ray shell. In contrast, the western side, where the gamma-ray significance is relatively low, shows slightly weaker CO intensity. These distributions of CO and H{\sc i} within $|b| < 1^\circ$ are consistent with those presented in Figure~B2 of Appendix B in \cite{2022A&A...666A.124A}.

Notably, we identify a molecular cloud that spatially coincides with the blowout-like structure in the gamma-ray emission. This feature was not reported in \cite{2022A&A...666A.124A}, because it lies outside the coverage of the Mopra CO survey data used in that study. The H{\sc i} clouds are mainly concentrated along the Galactic plane. Two bright H{\sc i} clouds, appearing white in Figure~\ref{fig2}(c), lie neatly along the outer boundaries of the northern and western parts of the gamma-ray shell. We also identify H{\sc i} dips along the shell at ($l$, $b$) $\sim$ (339\fdg7, $-$0\fdg9) and (339\fdg1, $-$0\fdg9).

Figures~\ref{fig2}(b) and \ref{fig2}(d) show the position--velocity ($p$--$v$) diagrams of CO and H{\sc i}. The $p$--$v$ diagram of CO reveals a cavity-like structure in the velocity range from $-$36 to $-$23~km~s$^{-1}$. Notably, the spatial extent of this CO cavity along the Galactic longitude is roughly consistent with the apparent diameter of the gamma-ray shell. Although the H{\sc i} $p$--$v$ diagram does not exhibit a similarly clear cavity-like structure, we find a large velocity gradient in H{\sc i} from ($V_\mathrm{LSR}$, $b$) $\sim$ ($-31$~km~s$^{-1}$, $0\fdg0$) to $\sim$ ($-27$~km~s$^{-1}$, $-1\fdg0$). The similar velocity gradient in H{\sc i} $p$--$v$ diagram was also observed in SNR RX~J1713.7$-$3946 \citep[see Figure 14b of][]{2012ApJ...746...82F}. This point is discussed in more detail in Section~\ref{Interstellar_gas}.

We also performed the same analysis in the velocity ranges of $V_\mathrm{LSR} = -60$--$-50$~km~s$^{-1}$ and $-48.5$--$-38.5$~km~s$^{-1}$, discussed by \red{\cite{2007A&A...468..993K} and} \cite{2022A&A...666A.124A}, with the results presented in Appendix~\ref{AppendixC}. As a result, we find no cavity-like structure in the $p$--$v$ diagram in either velocity range, comparable to that seen in Figure~\ref{fig2}(b) (see Appendix~\ref{AppendixC} for details).

\subsection{H{\sc i}-dip toward the southeastern gamma-ray peak} 
\label{hidip}
Figure~\ref{fig3}(a) shows an enlarged view of the southeastern part of the gamma-ray shell overlaid with the averaged H{\sc i} brightness temperature. We find that the H{\sc i} brightness decreases from 75~K to 63~K toward the gamma-ray peak of the southeastern shell at ($l$, $b$) $\sim$ (339\fdg1, $-$0\fdg9). The H{\sc i} depression (hereafter H{\sc i}-dip) shows a good spatial correspondence with the gamma-ray peak within the H.E.S.S. PSF. 

Figure~\ref{fig3}(b) shows the line profiles of CO and H{\sc i} at $V_\mathrm{LSR}$ $\sim$$-$32~km~s$^{-1}$. The H{\sc i} brightness temperature exhibits a dip-like feature in the velocity range from $\sim$$-$40 to $\sim$$-$30~km~s$^{-1}$, corresponding to the CO emission peaked at $V_\mathrm{LSR}$ $\sim$$-32$~km~s$^{-1}$. This point will be discussed in detail in Section~\ref{total_ism} and Appendix~\ref{AppendixB}.

We also note the presence of another H{\sc i}-dip at ($l$, $b$) $\sim$ (339\fdg75, $-$0\fdg90), which exhibits CO and H{\sc i} spectra similar to those described above. However, we do not discuss this feature further, as it is located just outside the gamma-ray shell.

\begin{figure*}[]
\begin{center}
\includegraphics[width=\linewidth,clip]{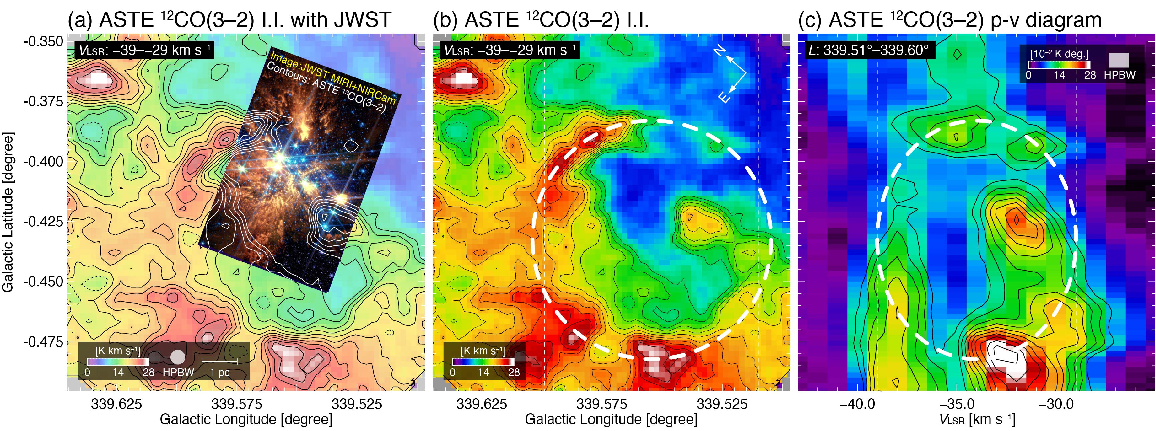}
\caption{Enlarged view of Westerlund~1 and its surroundings. (a, b) Integrated intensity maps of the ASTE $^{12}$CO($J$~=~3--2). The integration velocity range is from $-39$ to $-29$~km~s$^{-1}$. The lowest contour level and the interval are 12.0 and 1.5 K~km~s$^{-1}$, respectively. The background colored image in (a) represents the JWST MIRI/NIRCam composite image \citep{2025A&A...693A.120G}. The red, green, and blue represent the F1130W, F770W, and F444W filters. (c) Position--velocity diagram of the ASTE $^{12}$CO($J$~=~3--2). The integration range is from 339\fdg51 to 339\fdg60 in Galactic Longitude. The lowest contour level and the interval are 0.1 and 0.02 K~degree, respectively. The dashed circle in (b) and (c) indicates boundaries of the CO expanding shell (see the text).}
\label{fig5}
\end{center}
\end{figure*}%

\begin{figure*}[]
\begin{center}
\includegraphics[width=\linewidth,clip]{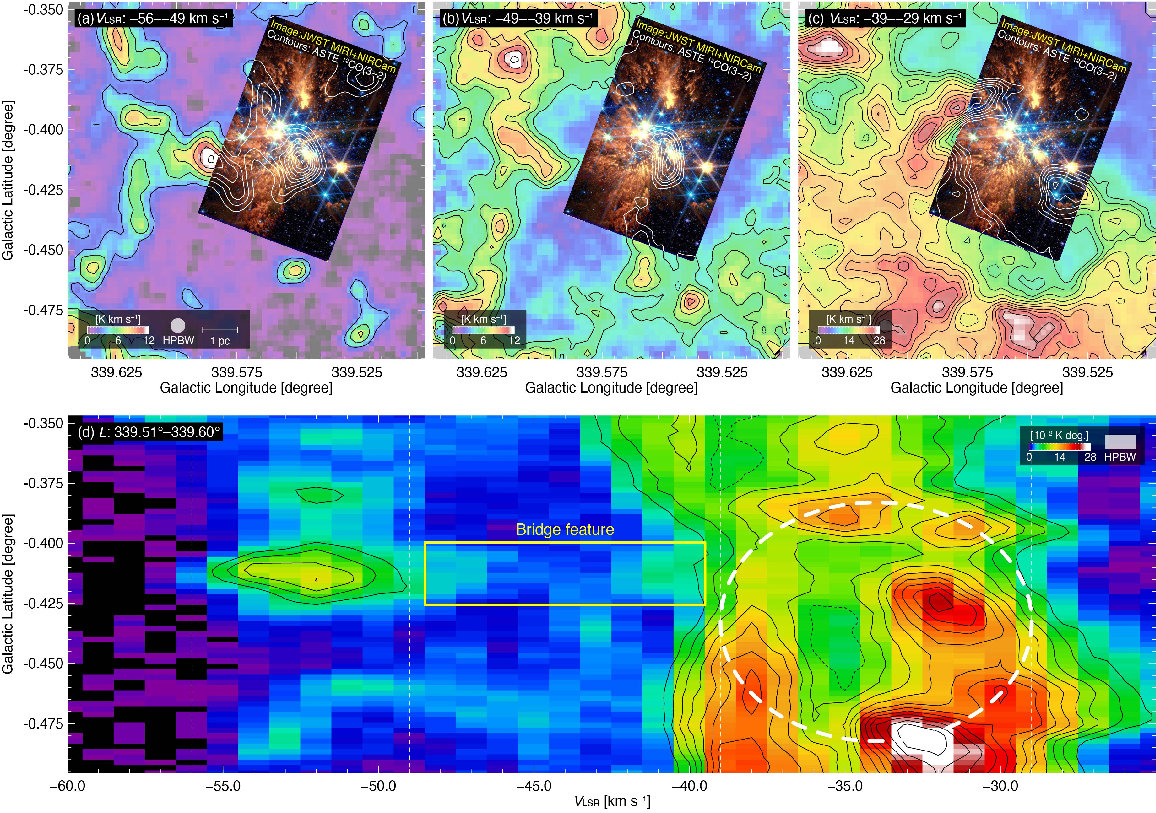}
\caption{(a--c) Same integrated intensity maps as in Figure~\ref{fig5}(a), but the intergrated velocity range is from $-56$ to $-49$~km~s$^{-1}$ for (a) and $-49$ to $-39$~km~s$^{-1}$ for (b). The contour levels are from 3, 5, 7, 9, 11, 13, 15, 17, and 19 K~km~s$^{-1}$ for (a) and from 4.8, 6.0, 7.2, 8.4, 9.6, 10.8, and 12.0 K~km~s$^{-1}$ for (b). (d) Same position--velocity diagram as in Figure~\ref{fig5}(c), but the displayed velocity range is from $-60$ to $-25$~km~s$^{-1}$. The integration range is the same as in Figure~\ref{fig5}(c). The lowest contour level and the interval are 0.05 and 0.02~K~degree, respectively. The yellow box and dashed circle indicate the bridge feature and the expanding shell (see the text).}
\label{fig6}
\end{center}
\end{figure*}%

\subsection{Detailed CO distribution in the vicinity of Wd1}
\label{aste}
To examine the possible association between the interstellar gas at $V_{\mathrm{LSR}} \sim$$-32$~km~s$^{-1}$ and the Wd1 cluster, we focus on the region immediately surrounding Wd1.

Figures~\ref{fig5}(a) and \ref{fig5}(b) show the velocity-integrated intensity maps of ASTE $^{12}$CO($J$~=~3--2) in the vicinity of the Wd1 cluster. A comparison with the JWST MIRI/NIRCam composite image overlaid in Figure~\ref{fig5}(a) reveals that CO clouds at $V_\mathrm{LSR} = -39$--$-29$~km~s$^{-1}$ are located to the north and south of the cluster. Most importantly, as indicated by the white dashed circle in Figure~\ref{fig5}(b), a CO cavity with a radius of $\sim$$3.4$~pc is clearly identified around the Wd1 cluster.

Figure~\ref{fig5}(c) shows the $p$--$v$ diagram of $^{12}$CO($J$~=~3--2) toward the Wd1 cluster. We newly identify a small cavity-like structure of CO, whose velocity extent is $\sim$10~km~s$^{-1}$ and a diameter in Galactic latitude is comparable to that of the cavity-like structure of CO shown in Figure~\ref{fig5}(b).

Figures~\ref{fig6}(a)--\ref{fig6}(c) show the same velocity-integrated maps of $^{12}$CO($J$~=~3–2), but for different velocity \red{ranges. The two clouds identified in the ASTE data at $V_{\mathrm{LSR}} = -56$--$-49$~km~s$^{-1}$ and $-49$--$-39$~km~s$^{-1}$} are located toward the center of the Wd1 cluster and are spatially coincident with the infrared dark lane seen in the JWST image. Interestingly, these two clouds and the cloud at $V_{\mathrm{LSR}} = -39$--$-29$~km~s$^{-1}$ appear to exhibit a complementary spatial distribution.

Figure~\ref{fig6}(d) shows the same $p$--$v$ diagram as in Figure~\ref{fig5}(c), but over a broader velocity range of $-60$ to $-25$ km s$^{-1}$. In velocity space, the cloud at $V_{\mathrm{LSR}} = -56$--$-49$~km~s$^{-1}$ corresponds to a decrease in intensity of the $V_{\mathrm{LSR}} = -39$--$-29$~km~s$^{-1}$ component at the same Galactic latitude, suggesting a complementary distribution. We also identify a velocity component connecting the clouds at $V_{\mathrm{LSR}} = -56$--$-49$~km~s$^{-1}$ and $V_{\mathrm{LSR}} = -39$--$-29$~km~s$^{-1}$ (hereafter, the bridge feature). This component lies at intermediate velocities between the two clouds.

\section{Discussion}
\label{discussion}
\subsection{Interstellar gas associated with HESS~J1646$-$458 and the Wd1 cluster}
\label{Interstellar_gas}
The previous CO and H{\sc i} studies reported no interstellar gas spatially coinciding with Wd1 or the gamma-ray peaks of HESS~J1646$-$458 \citep{2010ApJ...713L..45L,2012A&A...537A.114A,2022A&A...666A.124A}. In this section, we argue that the CO and H{\sc i} clouds at $V_\mathrm{LSR}$ $\sim$$-36$--$-23$~km~s$^{-1}$ are physically associated with both Wd1 and HESS~J1646$-$458. In addition, we discuss the possibility that a cloud at $V_\mathrm{LSR} \sim$$-55$~km~s$^{-1}$ is also physically associated with the Wd1 cluster.

\subsubsection{Warm CO clouds associated with the Wd1 cluster}
We first note that the CO cloud at $V_\mathrm{LSR}$ $\sim$$-$32~km~s$^{-1}$ may be located at a similar distance to Wd1. This interpretation is motivated by the association of a CO cloud with an arc-like feature seen at 8~$\mu$m, possibly suggesting that PAH dust within the molecular cloud is excited by strong FUV radiation from Wd1 (Figure~\ref{fig1}d). Several H{\sc ii} regions have been identified within $\sim$$2^\circ$ of Wd1, and illumination by FUV radiation from other massive stars cannot be completely ruled out \citep[e.g.,][]{2007A&A...468..993K}. However, no H{\sc ii} region capable of producing a PDR of this scale is found in the opening direction of the arc-like feature other than Wd1.

More direct evidence supporting this interpretation comes from a comparison between the JWST image and the ASTE $^{12}$CO($J$~=~3--2) data in the central region of Wd1 (Figure~\ref{fig5}a). The newly identified warm molecular cloud spatially corresponds to structures seen in the JWST image. In particular, the cavity-like distribution with a diameter of $\sim$$6.8$~pc seen in Figure~\ref{fig5}(b) is reminiscent of wind-blown bubbles observed around Wolf-Rayet stars \citep[e.g.,][]{2025PASA...42..101B}. If this interpretation is correct, the cavity-like structure with a similar spatial extent seen in the $p$--$v$ diagram (Figure~\ref{fig5}c) indicates expanding gas motion with a velocity of $\Delta V \sim$$5$~km~s$^{-1}$. The corresponding dynamical timescale of the expanding cloud is estimated to be $\sim$$0.7$~Myr. Given the estimated age of the Wd1 cluster \citep[$\sim$3--5~Myrs, e.g.,][]{2005A&A...434..949C,2006MNRAS.372.1407C,2008A&A...478..137B}, this structure is likely a relatively recently formed wind-blown bubble. These results strongly support a physical association between the CO clouds at $V_\mathrm{LSR} \sim$$-32$~km~s$^{-1}$ and the Wd1 cluster.

\subsubsection{Cloud-cloud collisions toward Wd1}
We next argue that the CO cloud at $V_{\mathrm{LSR}} = -56$--$-49$~km~s$^{-1}$, detected with ASTE toward Wd1, is also physically associated with the Wd1 cluster. Specifically, we consider the possibility that this cloud represents a collision partner of the $V_{\mathrm{LSR}} \sim$$-32$~km~s$^{-1}$ cloud and may have contributed to the formation of Wd1 through a  ``cloud-cloud collision.''

As reviewed by \citet{2021PASJ...73S...1F}, many super star clusters, including Wd1, and high-mass star(s) are thought to be formed via strong gas compression induced by collisions between two individual clouds. The observational signatures of such cloud-cloud collisions typically include (1) a supersonic velocity separation between two molecular clouds, (2) a complementary spatial distribution resembling a key-and-keyhole morphology, and (3) the presence of an intermediate-velocity component ``a bridge feature'' connecting the two clouds in velocity space.

\begin{figure*}[]
\begin{center}
\includegraphics[width=\linewidth,clip]{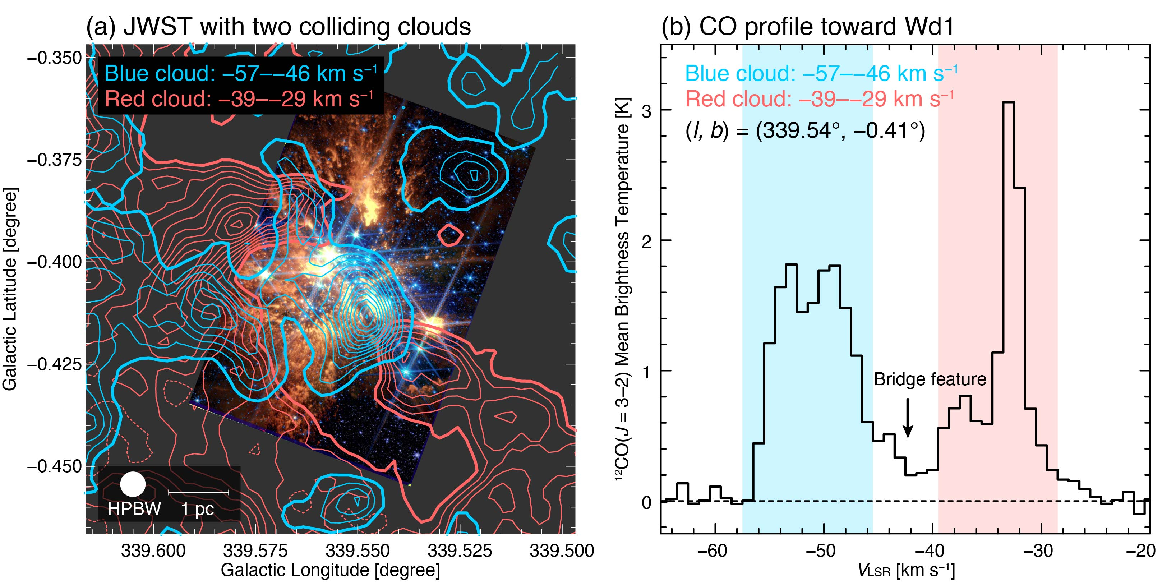}
\caption{(a) ASTE $^{12}$CO($J$~=~3--2) distributions of the blue cloud ($V_\mathrm{LSR}$: $-57$--$-46$~km~s$^{-1}$) and the red cloud ($V_\mathrm{LSR}$: $-39$--$-29$~km~s$^{-1}$) superimposed on the JWST image in Figure~\ref{fig5}(a). The lowest contour levels and the interval are 3 and 2~K~km~s$^{-1}$ for the blue cloud and are 12 and 1.5~K~km~s$^{-1}$ for the red cloud, respectively. (b) The spatially averaged $^{12}$CO($J$~=~3--2) profile around at ($l$, $b$) $\sim$ (339\fdg54, $-0\fdg41$).}
\label{fig7}
\end{center}
\end{figure*}%

As shown in Figures~\ref{fig6}(a) and \ref{fig6}(c), the $V_{\mathrm{LSR}} = -56$--$-49$~km~s$^{-1}$ cloud is located toward the central region of Wd1 and exhibits a complementary spatial distribution with the $V_{\mathrm{LSR}} \sim$$-32$ km s$^{-1}$ cloud, which surrounds the cluster. The velocity separation between the two clouds is approximately 20~km~s$^{-1}$, comparable to that observed in similar super star cluster Westerlund~2 \citep[e.g.,][]{2009ApJ...696L.115F,2010ApJ...709..975O}. Furthermore, the $V_{\mathrm{LSR}} = -49$--$-39$~km~s$^{-1}$ component shows a bridge feature that spatially and kinematically connects the two clouds (Figures~\ref{fig6}b and \ref{fig6}d). These lines of evidence support the interpretation that the $V_{\mathrm{LSR}} \sim$$-55$ km s$^{-1}$ cloud is also physically associated with Wd1. In other words, the two molecular clouds, despite their different line-of-sight velocities, may coexist at a common distance of $\sim$$3.9$~kpc.

Figure~\ref{fig7}(a) presents an overall view of the two molecular clouds associated with Wd1. The velocity range of $V_{\mathrm{LSR}} = -57$--$-46$~km~s$^{-1}$ corresponds to the blue-shifted cloud (hereafter, the blue cloud), while $V_{\mathrm{LSR}} = -39$--$-29$~km~s$^{-1}$ corresponds to the red-shifted cloud (hereafter, the red cloud). In this configuration, the blue cloud is located in front of the cluster. The intensity peak of the blue cloud coincides with a faint infrared dark lane seen in the JWST image, supporting this relative geometry.

Figure~\ref{fig7}(b) shows the CO spectrum averaged over the spatially overlapping region of the two clouds. A bridge feature connecting the two velocity components separated by $\sim$$20$~km~s$^{-1}$ is clearly identified. Since this study relies on a single $^{12}$CO($J$~=~3--2) transition, a quantitative evaluation of the physical conditions of colliding clouds---such as kinetic temperature and number density of molecular hydrogen---and a detailed test of the cloud-cloud collision scenario are beyond the scope of this paper. Accepted ALMA Band~6 observations (PI: H. Sano; \#2025.1.00296.S) will enable us to perform CO multi-$J$ line analysis to further investigate the physical association of the two clouds in the near future.

We here argue that the cloud-cloud collision scenario and the physical association of the two clouds are also consistent with several previous observations toward Wd1. First, the coexistence of clouds with different LSR velocities at a common distance suggests that kinematic distance estimates in this region may be unreliable. One of the clouds may represent the systemic velocity component of Wd1, although we do not pursue this issue further here.

The H{\sc i} bubbles reported by \citet{2007A&A...468..993K} may also be interpreted as structures influenced by Wd1 if the $V_{\mathrm{LSR}} \sim$$-55$~km~s$^{-1}$ cloud is indeed one of the colliding components. In addition, the stellar LSR velocity \citep[$-43 \pm 4$~km~s$^{-1}$,][]{2022A&A...664A.146N} and other kinematic tracers of H{\sc ii} regions and masers in Wd1 \citep[$\sim$$-50$--$-38$~km~s$^{-1}$,][]{2003A&A...397..133R,2012ApJ...760...65F} can be naturally explained as an intermediate velocity between the two colliding clouds. Magnetohydrodynamic simulations further suggest that dense cores formed through cloud-cloud collisions do not necessarily retain the original velocities of the parent clouds, supporting this interpretation \citep[e.g.,][]{2018PASJ...70S..53I}.

Taken together, these lines of evidence support the scenario in which both the $V_{\mathrm{LSR}} \sim$$-32$~km~s$^{-1}$ and $-55$~km~s$^{-1}$ clouds are physically associated with Wd1 and represent remnants of its natal molecular clouds. Among the proposed interpretations, this scenario provides the most coherent explanation of the multiwavelength observational properties of Wd1 reported to date.

\subsubsection{CO/H{\sc i} clouds associated with HESS~J1646$-$458}
We next discuss the possibility that these molecular clouds are also associated with HESS~J1646$-$458. In particular, the spatial correspondence between the molecular clouds and the TeV gamma-ray shell supports this association (Figure~\ref{fig2}a), especially given that previous studies suggested a hadronic origin for the gamma-ray emission \citep[e.g.,][]{2022A&A...666A.124A}. Notably, identifying a molecular cloud corresponding to the tail-like structure of the gamma-ray emission represents a new result made possible by our extended spatial coverage of CO, which was not achievable in earlier studies.

The H{\sc i} cloud at the same LSR velocity is also likely to contribute to the gamma-ray emission. This is supported by the H{\sc i}-dip corresponding to the peak of the southeastern gamma-ray shell, which is \red{possibly} due to H{\sc i} self-absorption. In this scenario, the spin temperature of H{\sc i} is reduced due to an increase in gas density, and the observed H{\sc i} intensity appears suppressed according to the radiative transfer equation when a warm H{\sc i} cloud (expected profile in Figure~\ref{fig3}b) is present behind the self-absorbing gas \citep[e.g.,][]{1978AJ.....83.1607S}. A similar H{\sc i} feature has been observed in other gamma-ray SNRs, consistent with the idea that dense and cold H{\sc i} can serve as a target for CR protons \citep[e.g.,][]{2012ApJ...746...82F,2014ApJ...788...94F}.

Finally, we argue that the cavity-like structure seen in the CO $p$--$v$ diagram provides a key link between Wd1, HESS~J1646$-$458, and the interstellar gas at $V_\mathrm{LSR}$ $\sim$$-36$--$-23$~km~s$^{-1}$ because strong stellar winds from Wd1 likely created this cavity. Here, we can derive the systemic velocity to be $V_\mathrm{LSR}$ $\sim$$-30 \pm 3$~km~s$^{-1}$, and its expansion velocity of $\Delta V$ $\sim$6.5~km~s$^{-1}$. By assuming the source distance of 3.9~kpc, the dynamical time scale of the expanding gaseous shell is to be $\sim$5.7~Myr, which is roughly consistent with the previous age estimation of Wd1 \citep[$\sim$3--5~Myrs, e.g.,][]{2005A&A...434..949C,2006MNRAS.372.1407C,2008A&A...478..137B}. Moreover, the H{\sc i} clouds distributed along the northern and eastern parts of the TeV gamma-ray shell (Figure~\ref{fig2}c), as well as the large velocity gradient seen in the H{\sc i} $p$--$v$ diagram (Figure~\ref{fig2}d), closely resemble those observed in RX~J1713.7$-$3946, where similar expanding gaseous motion has been reported \citep[see Figure~14b of][]{2012ApJ...746...82F}. Such a velocity gradient is expected when cold H{\sc i} gas associated with the foreground side of an expanding shell produces self-absorption, and is therefore consistent with the interpretation discussed above.

In light of these considerations, we conclude that the CO and H{\sc i} clouds at $V_\mathrm{LSR}$ $\sim$$-36$--$-23$~km~s$^{-1}$ are physically associated with both Wd1 and HESS~J1646$-$458. In addition, on smaller scales toward the central region of Wd1, we conclude that the $V_\mathrm{LSR} \sim$$-55$~km~s$^{-1}$ cloud is also physically associated with the Wd1 cluster. In the following discussion, we focus on the $V_\mathrm{LSR} \sim$$-36$--$-23$~km~s$^{-1}$ range in order to compare the large-scale gas distribution with the gamma-ray emission associated with HESS~J1646$-$458\red{\footnote{\red{The CO cloud at $V_\mathrm{LSR} \sim$$-55$~km~s$^{-1}$ is associated with the central region of the Wd1 cluster on a small angular scale ($\sim$0\fdg1), but shows neither expanding motion nor spatial correspondence with the gamma-ray shell on larger scales ($\sim$2$^\circ$) in the CO or H{\sc i} data. This compact distribution may be consistent with localized cloud--cloud collision related to the formation of the cluster. We therefore consider that this component does not contribute significantly to the large-scale gamma-ray emission of HESS~J1646$-$458.}}}.

\subsection{Origin of Gamma-rays}
\label{origin}
\cite{2022A&A...666A.124A} have suggested that Wd1 is the most plausible source of the bulk of the gamma-ray emission from HESS~J1646$-$458. They first ruled out other objects in the vicinity that are unrelated to Wd1 (the low-mass X-ray binary 4U~1642$-$45, PSR~J1648$-$4611, and PSR~J1650$-$4601) as incapable of accounting for the spatially extended and energetically broad gamma-ray emission of HESS~J1646$-$458. The authors argued that leptonic scenarios are disfavored, based on the lack of a gamma-ray peak toward Wd1, the energy-independent morphology, the large spatial extent of HESS~J1646$-$458, and the absence of multi-wavelength signatures of high-energy electrons. However, viable leptonic models have also been proposed \citep[e.g.,][]{2023A&A...671A...4H}, indicating that the emission mechanism remains under discussion. In this context, our MeerKAT analysis and recent eROSITA results \red{\citep{2025A&A...695A...3H}} do not reveal clear evidence of diffuse radio synchrotron emission associated with the gamma-ray shell\footnote{The 1.3~GHz radio continuum data used in this study were obtained with MeerKAT alone, without combination with single-dish data \citep[see][]{2024MNRAS.531..649G}. As a result, the dataset may lack sensitivity to extended emission and suffer from the negative bowl. In either case, further follow-up studies will be essential to better understand the nature of the radio continuum emission toward HESS~J1646$-$458.} (see Figure~\ref{fig1}b of this paper). While these constraints do not uniquely determine the emission mechanism, they are consistent with a scenario in which hadronic emission plays a significant role.

Second, they pointed out that at least one supernova explosion must have occurred in the cluster, given the presence of the magnetar CXOU~J164710.2$-$455216 within Wd1. This, together with the strong stellar winds from the cluster, could plausibly enhance particle acceleration and make a hadronic origin viable. Nevertheless, assuming a target gas density of $n_\mathrm{p}$ $\sim$5~cm$^{-3}$, they noted that explaining the observed gamma-ray luminosity would require more than ten supernovae. Although the absence of CO or H{\sc i} gas corresponding to the gamma-ray emission may present a challenge for the hadronic scenario, it has been argued that this could be due to uncertainties in the distribution of the target gas. 

In addition, when accounting for CR diffusion, the observed gamma-rays would imply the presence of CR protons with energies up to several PeV confined within Wd1. Second-order Fermi acceleration through strong magnetohydrodynamic turbulence driven by multiple supernovae has also been considered; however, the relatively small size of HESS~J1646$-$458 compared to what is typically expected for superbubbles poses a challenge to this interpretation. Based on these considerations, the authors concluded that the most promising acceleration mechanism is the termination shock of the collective cluster wind.

In all hadronic scenarios, the presence of interstellar protons spatially corresponding to the gamma-ray emission is a necessary condition. In this section, we quantitatively examine the spatial relation between the gamma-rays and the interstellar protons, providing strong observational support for the hadronic interpretation proposed by \cite{2022A&A...666A.124A}.

\begin{figure*}[]
\begin{center}
\includegraphics[width=\linewidth,clip]{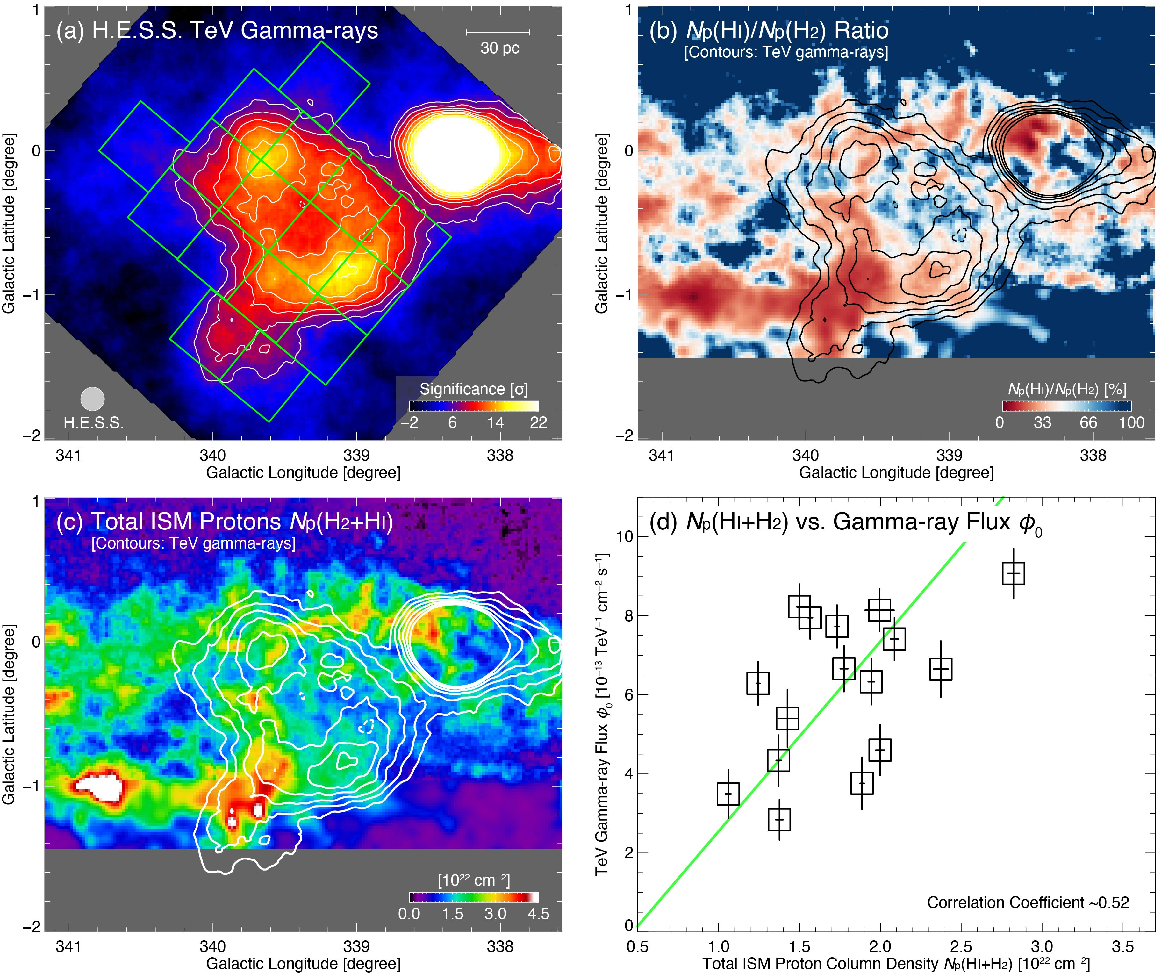}
\caption{(a) Same as Figure~\ref{fig1}a, but overlaid with 16 boxes that are used for the gamma-ray flux $\phi_\mathrm{0}$ calculation in \cite{2022A&A...666A.124A}. (b) Ratio map of $N_\mathrm{p}$(H{\sc i})/$N_\mathrm{p}$(H$_2$). (c) Distribution of the total ISM proton column density $N_\mathrm{p}$(H{\sc i} + H$_2$). The superposed contours are the same as shown in Figure~\ref{fig1}a. (d) Scatter plot between $N_\mathrm{p}$(H{\sc i} + H$_2$) and $\phi_\mathrm{0}$. The green line shows the linear regression by $\chi^2$ fitting (see the text). The calculated correlation coefficient is $\sim$0.52.}
\label{fig4}
\end{center}
\end{figure*}%

\subsubsection{Comparison with Total Interstellar Protons}
\label{total_ism}
To obtain the distribution of total interstellar protons with HESS~J1646$-$458, we estimate proton column densities comprising both the molecular and atomic components \citep[e.g.,][]{2012ApJ...746...82F,2019ApJ...876...37S}. The interstellar proton column density of the molecular component $N_\mathrm{p}$(H$_2$) can be calculated from the followings:
\begin{eqnarray}
N_\mathrm{p}(\mathrm{H_2}) &=& 2 \times N(\mathrm{H_2}) \; \mathrm{(cm^{-2})},\\
&=& 2 \times X_\mathrm{CO} \cdot W(\mathrm{CO}) \; \mathrm{(cm^{-2})}.
\label{eq1}
\end{eqnarray} 
where $N$(H$_\mathrm{2}$) is a molecular hydrogen column density, $X_\mathrm{CO}$ is a CO-to-H$_2$ conversion factor, and $W$(CO) is the velocity-integrated intensity of $^{12}$CO($J$~=~1--0). Here, we used $X_\mathrm{CO} = 2 \times 10^{20}$ (K~km~s$^{-1}$)$^{-1}$~cm$^{-2}$ \citep{2013ARA&A..51..207B}, and the integrated velocity range is from $-36$ to $-23$~km~s$^{-1}$.

As for the interstellar proton column density of the atomic component, $N_\mathrm{p}$(H{\sc i}), the optical depth of H{\sc i} must be carefully considered. \cite{2012ApJ...746...82F} performed a correction for optically thick H{\sc i} by analyzing H{\sc i} self-absorption dips using the radiative transfer equation, such as the one shown in Figure~\ref{fig3}b. They found that the optical depth of H{\sc i}-dip associated with CO emission is typically around 0.7, resulting in an increase of $N_\mathrm{p}$(H{\sc i}) by a factor of $\sim$1.2 compared to estimates assuming optically thin conditions \citep[optical depth $\ll 1$, c.f. ][]{1990ARA&A..28..215D} as below:
\begin{eqnarray}
N_\mathrm{p}(\mathrm{H}{\textsc{i}}) = 1.823 \times 10^{18}  \cdot W(\mathrm{H}{\textsc{i}}) \;(\mathrm{cm}^{-2}),  
\label{eq2}
\end{eqnarray} 
where $W$(H{\sc i}) is the velocity-integrated intensity of H{\sc i}. In the case of HESS~J1646$-$458 as shown in Figure~\ref{fig3}, we calculated the optical-depth corrected $N_\mathrm{p}$(H{\sc i}) to be $\sim$2.0--$2.1 \times 10^{21}$~cm$^{-2}$ that is typically $\sim$1.3 times higher than the optically thin assumption (see Appendix~\ref{AppendixB}).

Later, \cite{2017ApJ...850...71F} derived an empirical method to correct the optical depth of H{\sc i} by using dust opacity distributions obtained from Planck and IRAS \citep{2014A&A...571A..11P}, taking into account the nonlinear properties of dust \citep[][]{2015ApJ...798....6F,2013ApJ...763...55R,2017ApJ...838..132O}. This approach enables us to convert from $W$(H{\sc i}) to an optical-depth corrected $N_\mathrm{p}$(H{\sc i}). In the present study, we adopt the method of \cite{2017ApJ...850...71F} to derive the map of $N_\mathrm{p}$(H{\sc i}).

Figure~\ref{fig4} shows comparisons among the gamma-ray significance, fluxes, and the interstellar proton column density. The ratio map of $N_\mathrm{p}$(H{\sc i})/$N_\mathrm{p}$(H$_2$) suggests that \red{the interstellar protons} toward the gamma-ray shell of HESS~J1646$-$458 are \red{predominantly associated with} molecular \red{gas} (Figure~\ref{fig4}b). The distribution of the total interstellar proton column density $N_\mathrm{p}$(H{\sc i} $+$ H$_\mathrm{2}$), as shown in Figure~\ref{fig4}c, generally follows the spatial distribution of CO and shows good spatial correspondence with the gamma-ray significance contours. 

To quantitatively investigate this relation, we compared the gamma-ray fluxes obtained for the 16 regions \citep[][green boxes in Figure~\ref{fig4}a]{2022A&A...666A.124A} with the total interstellar proton column densities newly estimated in this study. Figure~\ref{fig4}d shows a scatter plot between the total interstellar proton column density $N_\mathrm{p}$(H{\sc i} $+$ H$_\mathrm{2}$) and TeV gamma-ray flux $\phi_\mathrm{0}$. We find a moderate correlation with a correlation coefficient of $\sim$0.52, which was fitted by the least-squares method using the \texttt{MPFITEXY} routine \citep{2010MNRAS.409.1330W}. This is particularly intriguing given that the analysis regions include not only the interior of the gamma-ray shell but also its fainter outer parts. In the case of hadronic gamma-rays, if the spatial distribution of accelerated CR protons is uniform, the resulting gamma-ray flux should be proportional to the target interstellar proton density. However, when including regions outside the shell, the CR distribution may no longer be approximated as uniform due to energy- and time-dependent diffusion effects. Nevertheless, the fact that the spatial correlation is still observed suggests that the CR diffusion time may be sufficiently long for the CR protons to effectively interact with the interstellar protons and produce hadronic gamma-rays. In any case, the good spatial correspondence between the gamma-ray flux and the total interstellar proton density obtained in this study strongly supports the hadronic origin of the gamma-rays proposed by \cite{2022A&A...666A.124A}.

\subsubsection{Total Energy of Accelerated CR Protons}
As discussed in Sections~\ref{origin} and \ref{total_ism}, the gamma-ray emission from HESS~J1646$-$458 is consistent with being primarily produced by a hadronic process. In such a scenario, the total energy of accelerated CR protons, $W_\mathrm{p}$, is known to be proportional to the gamma-ray luminosity, $L_\gamma$, and inversely proportional to the target gas density, $n_\mathrm{p}$. In this section, we estimate the total mass and average density of interstellar protons associated with HESS~J1646$-$458, and derive $W_\mathrm{p}$ accordingly. 

To derive the $W_\mathrm{p}$ value of HESS~J1646$-$458, we first estimate the total mass and averaged density of interstellar protons associated with HESS~J1646$-$458. The mass of molecular clouds $M_\mathrm{co}$ and atomic gas $M_\mathrm{HI}$ can be estimated using the following equations:
\begin{eqnarray}
M_\mathrm{CO} &=& m_{\mathrm{p}} \mu \Omega D^2 \sum_{i} N_i(\mathrm{H}_2),\\
M_\mathrm{HI} &=& m_{\mathrm{p}} \Omega D^2 \sum_{i} N_i(\mathrm{H{\textsc{i}}}),
\label{eq3}
\end{eqnarray}
where $m_\mathrm{p}$ is the mass of an interstellar proton, $\mu = 2.8$ is the mean molecular weight, $\Omega$ is the solid angle of each data pixel, and $D$ is the distance to HESS~J1646$-$458. By adopting $D = 3.9$~kpc, we estimated the masses of the molecular and atomic gas within the region enclosed by the 6$\sigma$ contour of the TeV gamma-ray significance map to be $M_\mathrm{CO}$ $\sim$$1.2 \times 10^6~M_{\odot}$ and $M_\mathrm{HI}$ $\sim$$0.4 \times 10^6~M_{\odot}$, respectively. The total gas mass is, therefore, estimated to be $\sim$$1.6 \times 10^6~M_{\odot}$. The average target gas densities of molecular and atomic gas, $n_\mathrm{p}$(H$_2$) and $n_\mathrm{p}$(H{\sc i}), within the same region were estimated to be $n_\mathrm{p}$(H$_2$) $\sim$80~cm$^{-3}$ and $n_\mathrm{p}$(H{\sc i}) $\sim$30~cm$^{-3}$, respectively, under the assumption that all interstellar protons are uniformly distributed within the effective spherical volume enclosed by the 6$\sigma$ contour. Accordingly, the total interstellar proton density was estimated to be $n_\mathrm{p}$ $\sim$110~cm$^{-3}$.

The total energy of accelerated CR protons $W_\mathrm{p}$ could be explained by \cite{2022A&A...666A.124A} as below:
\begin{eqnarray}
W_\mathrm{p} \sim 6 \times 10^{51} (d / 3.9~\mathrm{kpc})^2 (n_\mathrm{p} / 1~\mathrm{cm}^{-3})^{-1} \;(\mathrm{erg}).
\label{eq4}
\end{eqnarray}
By adopting $d = 3.9$~kpc and $n_\mathrm{p} = 110$~cm$^{-3}$, we finally derived $W_\mathrm{p}$ $\sim$$6\times 10^{49}$~erg. This value corresponds to only $\sim$6\% of the typical kinetic energy of a supernova explosion, which is on the order of $\sim$$10^{51}$~erg. If this scenario is correct, even a single supernova explosion that has already occurred within the Wd1 cluster, CXOU~J164710.2$-$455216, could potentially account for the required energy. It should be noted, however, that this does not rule out particle acceleration by the termination shock of the collective cluster wind from Wd1. \red{Recent theoretical studies have shown that cosmic rays accelerated at the termination shock within a superbubble can reproduce both the spectrum and spatial distribution of the gamma-ray emission from HESS~J1646$-$458 under a hadronic scenario \citep{2026JHEAp..5100560S}.}

In other words, the key observational result newly provided by this study is that the gamma-ray emission from HESS~J1646$-$458 is consistent with a hadronic origin through a detailed comparison with the CO and H{\sc i} datasets. Further investigation is essential to determine the underlying acceleration mechanism. On the observational side, further gamma-ray measurements with unprecedented sensitivity and angular resolution by the Cherenkov Telescope Array \citep[CTA,][]{2019scta.book.....C} will be crucial. On the theoretical side, more sophisticated models of particle acceleration and diffusion associated with cluster winds are required to enable meaningful comparisons with observations. In any case, as more gamma-ray sources continue to be detected from H{\sc ii} regions and stellar clusters \citep[e.g.,][]{2024NatAs...8..530P}, detailed investigations of the associated interstellar protons and high-resolution gamma-ray observations will be indispensable for revealing the full picture of CR proton acceleration in our Galaxy.

\section{Conclusions}
\label{conclusions}
We analyzed NANTEN/NANTEN2 $^{12}$CO($J$~=~1--0), ATCA \& Parkes archival H{\sc i}, newly obtained ASTE $^{12}$CO($J$~=~3--2) data, and other multiwavelength datasets to identify interstellar gas associated with the massive young stellar cluster Westerlund1 (Wd1) and the TeV gamma-ray source HESS~J1646$-$458. Our primary findings are summarized below.

\begin{enumerate}
\item We found molecular clouds at $V_\mathrm{LSR}\sim$$-32$~km~s$^{-1}$ that align with arc-like features in the Spitzer 8~$\mu$m image, likely tracing PAH emission enhanced by strong FUV radiation from massive stars of Wd1. At the same velocity range, ASTE $^{12}$CO($J$~=~3--2) observations reveal a cavity-like structure of molecular clouds with a diameter of $\sim$$6.8$~pc and an expansion velocity of $\sim$$5$~km~s$^{-1}$ toward the central region of Wd1. The corresponding dynamical timescale of $\sim$$0.7$~Myr suggests a relatively recent wind-blown bubble driven by the cluster.

\item The $V_\mathrm{LSR} \sim$$-32$ and $\sim$$-55$~km~s$^{-1}$ clouds identified by ASTE exhibit complementary spatial distributions and are connected by an intermediate-velocity bridge feature at $\sim$$-44$~km~s$^{-1}$ component. These characteristics are consistent with signatures of cloud-cloud collision reported in studies of triggered star formation toward super star clusters and high-mass stars. This \red{suggests} that both the $V_\mathrm{LSR} \sim$$-32$ and $\sim$$-55$~km~s$^{-1}$ clouds are physically linked to the Wd1 cluster and therefore coexist at a common distance of approximately 3.9~kpc despite their different LSR velocities.

\item The large-scale distributions of CO and H{\sc i} gas at $V_\mathrm{LSR}$ $\sim$$-36$--$-23$~km~s$^{-1}$ are in good agreement with the TeV gamma-ray shell of HESS~J1646$-$458. The position--velocity diagram of CO (possibly H{\sc i} as well) indicates the presence of an expanding gas motion, whose dynamical time scale is roughly consistent with the age of the Wd1 cluster. The H{\sc i} intensity depression toward the southern gamma-ray peak is likely caused by self-absorption of cold H{\sc i}, consistent with an expanding gaseous structure. This suggests that both molecular clouds and cold H{\sc i} gas act as targets of cosmic-ray protons to produce hadronic gamma-rays if the hadronic process is dominantly working.

\item The total mass of interstellar protons, consisting of neutral molecular and atomic hydrogen, associated with HESS~J1646$-$458 is estimated to be $\sim$$1.6 \times 10^6~M_\odot$. The moderate correlation between the total interstellar proton column density and gamma-ray flux provides alternative support for the hadronic origin by considering the absence of bright synchrotron emission and previous studies. 

\item The total energy of accelerated cosmic-ray protons is estimated to be $\sim$$6 \times 10^{49}$~erg, based on a target gas density of $\sim$110~cm$^{-3}$ and a source distance of $\sim$3.9~kpc. While a single supernova explosion may suffice to supply the required energy, this does not rule out efficient particle acceleration via the termination shock of the collective cluster wind from Wd1. Future high-sensitivity and high-angular-resolution observations of gamma-rays with the Cherenkov Telescope Array (CTA), alongside refined theoretical modeling and detailed studies of interstellar protons associated with other gamma-ray bright H{\sc ii} regions and stellar clusters, will be essential to elucidate the underlying acceleration mechanisms and evaluate their contribution to the population of Galactic cosmic-rays. 
\end{enumerate}

\begin{acknowledgments}
The NANTEN project is based on a mutual agreement between Nagoya University and the Carnegie Institution of Washington (CIW). We greatly appreciate the hospitality of all the staff members of the Las Campanas Observatory of CIW. NANTEN2 is an international collaboration of more than 10 universities and institutes: Nagoya University, Gifu University, Osaka Metropolitan University, University of Cologne, University of Bonn, Seoul National University, University of Chile, University of New South Wales, Macquarie University, University of Sydney, ETH Zurich, and the Nobeyama Radio Observatory. We are thankful to many Japanese public donors and companies who contributed to the realization of the project. The ASTE telescope is operated by the National Astronomical Observatory of Japan (NAOJ). Observations with ASTE were in part carried out remotely from Japan using NTT's GEMnet2 and its partner R\&E (Research and Education) networks, which are based on the AccessNova collaboration among the University of Chile, NTT Laboratories, and NAOJ. We thank Dr. Seiichi Sakamoto and Dr. Rei Enokiya for helping us with the ASTE operation. This work is based in part on observations made with the Spitzer Space Telescope, which was operated by the Jet Propulsion Laboratory, California Institute of Technology, under a contract with NASA. The MeerKAT telescope is operated by the South African Radio Astronomy Observatory, which is a facility of the National Research Foundation, an agency of the Department of Science and Innovation. The Australia Telescope Compact Array (ATCA) is part of the Australia Telescope National Facility (grid.421683.a), which is funded by the Australian Government for operation as a National Facility managed by CSIRO. We acknowledge the Gomeroi people as the traditional owners of the Observatory site. This work is based in part on observations made with the NASA/ESA/CSA James Webb Space Telescope (JWST). The data were obtained from the Mikulski Archive for Space Telescopes at the Space Telescope Science Institute, which is operated by the Association of Universities for Research in Astronomy, Inc., under NASA contract NAS 5-03127 for JWST. These observations are associated with program \#1905. Some of the data presented in this article were obtained from the Mikulski Archive for Space Telescopes (MAST) at the Space Telescope Science Institute. The specific observations analyzed can be accessed via \dataset[doi:10.17909/2732-nz41]{https://doi.org/10.17909/2732-nz41}. This work was also supported by JSPS KAKENHI grant Nos. 21H01136 (HS), 24H00246 (HS), 25K17435 (KT), and 26K21722 (HS). This work was supported by a University Research Support Grant from the National Astronomical Observatory of Japan (NAOJ). NAOJ ALMA Scientific Research Grant Code 2023-25A also supported this work. This research was supported by the grant of OML Project by the National Institutes of Natural Sciences (NINS program No. OML012309. This work was supported by the Tokai Pathways to Global Excellence (T-GEx), part of the MEXT Strategic Professional Development Program for Young Researchers, and by the establishment of university fellowships toward the creation of science technology innovation (Grant Number: JPMJFS2138). We thank the anonymous referee for valuable and constructive comments that significantly improved the manuscript and motivated additional analyses, including new CO observations of the central region of Wd1.
\end{acknowledgments}

\facilities{NANTEN, NANTEN2, ASTE, ATCA, MeerKAT, Spitzer, JWST, HESS}
\software{IDL Astronomy User's Library \citep{1993ASPC...52..246L}, MIRIAD \citep[][]{1995ASPC...77..433S}, \texttt{MPFITEXY} \citep{2010MNRAS.409.1330W}.}

\appendix

\section{Distributions of CO, H{\textsc{i}}, and $N_\mathrm{p}$(H{\textsc{i}} $+$ H$_\mathrm{2}$) for other velocity ranges}
\label{AppendixC}
\begin{figure*}[]
\begin{center}
\includegraphics[width=\linewidth,clip]{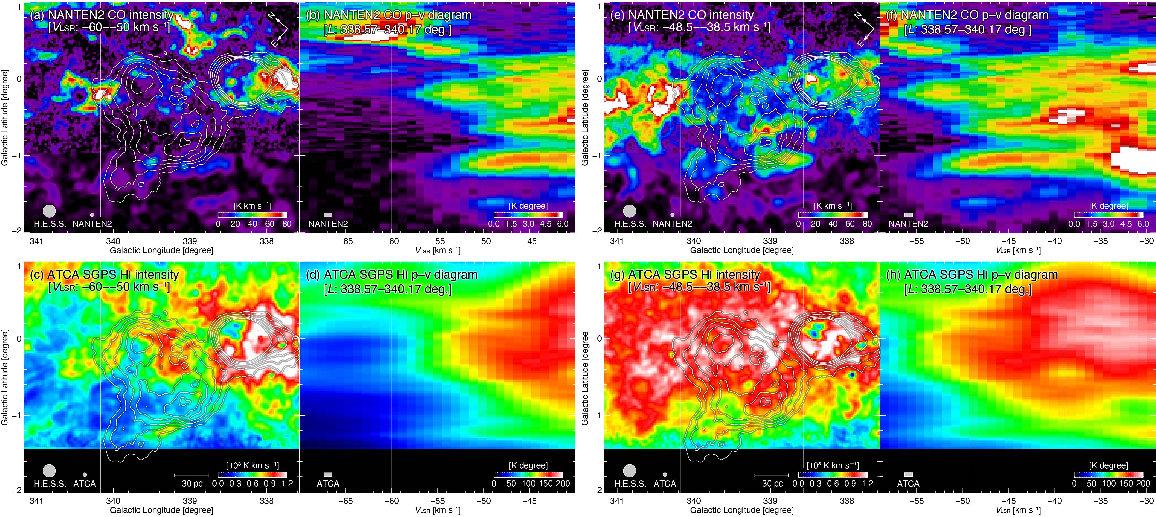}
\caption{Same integrated intensity maps and $p$--$v$ diagrams of CO and H{\sc i} as in Figure~\ref{fig2}, but the integration velocity range is from $-60$ to $-50$~km~s$^{-1}$ for (a--d) and from $-48.5$ to $-38.5$~km~s$^{-1}$ for (e--h). The superposed contours in each integrated intensity map and the integration range in Galactic Longitude for each $p$--$v$ diagram are the same as shown in Figure~\ref{fig2}.}
\label{figa1}
\end{center}
\end{figure*}

As a comparison with Figures~\ref{fig2}, we present CO and H{\sc i} integrated intensity maps and $p$--$v$ diagrams for the velocity ranges $V_\mathrm{LSR} = -60$--$-50$~km~s$^{-1}$ and $-48.5$--$-38.5$~km~s$^{-1}$ in Figures~\ref{figa1}(a--d) and \ref{figa1}(e--f), respectively. In contrast to the CO/H{\sc i} emission in the range $V_\mathrm{LSR} = -36$--$-23$~km~s$^{-1}$ as shown in Figure~\ref{fig2}, the CO distribution in these velocity intervals is sparse and clumpy, showing no clear spatial correspondence with the TeV gamma-ray shell. These spatial trends of CO/H{\sc i} are consistent with the previous studies \citep[e.g.,][]{2022A&A...666A.124A}. Moreover, the $p$--$v$ diagrams do not exhibit cavity-like structures whose diameters correspond to that of the gamma-ray shell.

\begin{figure*}[]
\begin{center}
\includegraphics[width=\linewidth,clip]{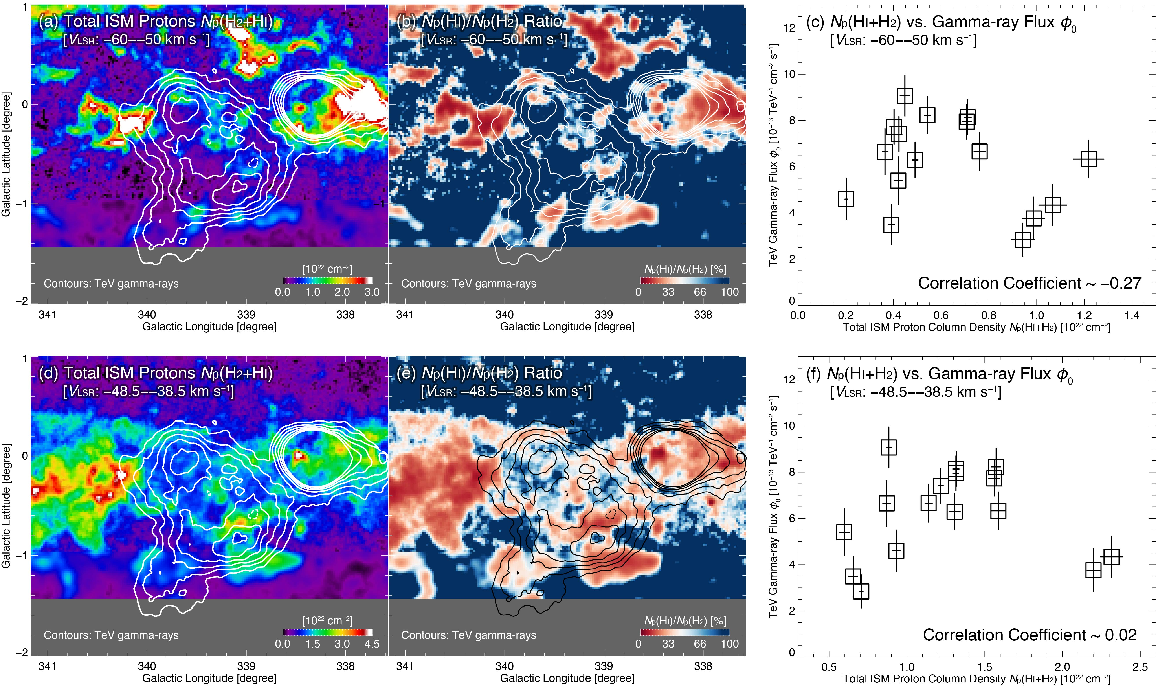}
\caption{Same maps of $N_\mathrm{p}$(H{\sc i} + H$_2$), $N_\mathrm{p}$(H{\sc i})/$N_\mathrm{p}$(H$_2$) ratio, and the scatter plots between $N_\mathrm{p}$(H{\sc i} + H$_2$) and $\phi_\mathrm{0}$ as in Figure~\ref{fig4}, but the velocity range of the ISM is from $-60$ to $-50$~km~s$^{-1}$ for (a--c) and from $-48.5$ to $-38.5$~km~s$^{-1}$ for (d--f).}
\label{figa2}
\end{center}
\end{figure*}

Figures~\ref{figa2}(a, b) and \ref{figa2}(d, e) show the spatial distributions of the total ISM proton column density, $N_\mathrm{p}$(H{\textsc{i}} $+$ H$_\mathrm{2}$), and the $N_\mathrm{p}$(H{\sc i})/$N_\mathrm{p}(\mathrm{H}_2)$ ratio for the velocity ranges $V_\mathrm{LSR} = -60$--$-50$~km~s$^{-1}$ and $-48.5$--$-38.5$~km~s$^{-1}$, respectively. The results of a quantitative comparison with the gamma-ray flux are presented in Figures~\ref{figa2}(c) and \ref{figa2}(f). These results indicate that $N_\mathrm{p}$(H{\textsc{i}} $+$ H$_\mathrm{2}$) tends to be lower toward regions of enhanced gamma-ray flux. The corresponding correlation coefficients between the total ISM proton column density and the gamma-ray flux are $\sim$$-0.27$ for the $V_\mathrm{LSR} = -60$--$-50$~km~s$^{-1}$ component and $\sim$$0.02$ for the $V_\mathrm{LSR} = -48.5$--$-38.5$~km~s$^{-1}$ component, indicating no significant correlation between the two quantities.

\section{Analysis of the H{\sc i} Self-Absorption Dip}
\label{AppendixB}
When estimating the column density of atomic hydrogen gas, $N_\mathrm{p}$(H{\sc i}), from the observed H{\sc i} brightness temperature, it is important to note that the emission intensity does not scale linearly with $N_\mathrm{p}$(H{\sc i}) if the optical depth effect is non-negligible. The observed H{\sc i} brightness temperature, $T_\mathrm{L}(v)$, as a function of line-of-sight velocity $v$, can be expressed using the radiative transfer equation as follows \citep[e.g.,][]{1978AJ.....83.1607S}:
\begin{eqnarray}
T_\mathrm{L}(v) = T_\mathrm{s}[1-e^{-\tau(v)}]+T^{\mathrm{FG}}_\mathrm{L}(v)+[T^{\mathrm{BG}}_\mathrm{L}(v)+T^{\mathrm{BG}}_\mathrm{C}]e^{-\tau (v)} - (T^{\mathrm{FG}}_\mathrm{C}+T^{\mathrm{BG}}_\mathrm{C}),
\label{eq5}
\end{eqnarray}
where $T_\mathrm{s}$, $\tau(v)$, $T^{\mathrm{FG}}_\mathrm{L}(v)$, and $T^{\mathrm{BG}}_\mathrm{L}(v)$ are the H{\sc i} spin temperature, the optical depth of H{\sc i}-dip, and the foreground (FG) and background (BG) H{\sc i} brightness temperature, respectively. Here, $T^{\mathrm{FG}}_\mathrm{C}$ and $T^{\mathrm{BG}}_\mathrm{C}$ represent the continuum brightness temperature at 1.4~GHz in the foreground and background of the H{\sc i} cloud. Since the radio continuum emission is weak in this direction (see Figure~\ref{fig1}), we ignore the $T^{\mathrm{FG}}_\mathrm{C}$ and $T^{\mathrm{BG}}_\mathrm{C}$ terms, following the method of \cite{2012ApJ...746...82F}. Therefore, if the integration range in LSR velocity is limited to that of the H{\sc i} cloud, the foreground H{\sc i} emission term, $T^{\mathrm{FG}}_\mathrm{L}(v)$, can also be neglected. In this case, the observed H{\sc i} intensity is determined solely by the spin temperature $T_\mathrm{s}$ and optical depth $\tau(v)$ of the H{\sc i} cloud, and the background H{\sc i} line emission.

To estimate the background component, $T^{\mathrm{BG}}_\mathrm{L}(v)$, we assume a linear interpolation across the H{\sc i} dip, as illustrated by the dashed line in the spectrum shown in Figure~\ref{fig3}(b).
For the spin temperature $T_\mathrm{s}$, we adopt three representative values, 10, 20, and 30~K, based on two considerations: (1) the peak brightness temperature of $^{12}$CO($J$~=~1--0) in the same velocity range is approximately $\sim$10~K, and (2) the H{\sc i} cloud is exposed to strong FUV radiation from Wd1, a superstar cluster. These values are motivated by the analysis of molecular clouds associated with another superstar cluster, as studied by \cite{2010ApJ...709..975O}. We then derive $\tau(v)$ for each assumed $T_\mathrm{s}$, and finally calculate the optical-depth corrected proton column density $N_\mathrm{p}$(H{\sc i}) using the following equation:
\begin{eqnarray}
N_{\mathrm{p}}(\mathrm{H}{\textsc{i}})=1.823\times10^{18} \int_{v_1}^{v_2} T_\mathrm{s}\: \tau(v)\: dv  \; \mathrm{(cm^{-2})}.
\label{eq6}
\end{eqnarray}
We adopt $v_1 = -36$~km~s$^{-1}$ and $v_2 = -23$~km~s$^{-1}$ as the velocity range of interest. Outside the velocity range corresponding to the H{\sc i}-dip, we assume $\tau(v) \ll 1$ and calculate $T_\mathrm{L}(v)$ using the approximation $T_\mathrm{L}(v) = T_\mathrm{s}\: \tau(v)$ (equation~\ref{eq2}). We then derived the maximum optical depth $\tau$ within this range to be approximately 0.3, 0.4, and 0.5 for $T_\mathrm{s} = 10$, 20, and 30~K, respectively. The optical-depth corrected H{\sc i} column density, $N_\mathrm{p}$(H{\sc i}), was estimated to be $\sim$2.0--$2.1 \times 10^{21}$~cm$^{-2}$, which is typically $\sim$1.3 times higher than the value derived under the optically thin assumption of H{\sc i}.

Toward the southeastern shell of HESS~J1646$-$458, $N_\mathrm{p}$(H$_2$) is as high as $\sim$$1.5 \times 10^{22}$~cm$^{-2}$. Therefore, even after correcting for self-absorption, the contribution of H{\sc i} gas to the total interstellar protons is relatively minor. This is because the interstellar protons associated with Wd1 and HESS~J1646$-$458 are predominantly in molecular form, H$_2$. However, in environments where atomic hydrogen dominates, or where H$_2$ and H{\sc i} are comparable in abundance (e.g., SNR RX~J1713.7-3946, \citeauthor{2012ApJ...746...82F} \citeyear{2012ApJ...746...82F}, \citeyear{2021ApJ...915...84F}; SNR~Vela~Jr., \citeauthor{2017ApJ...850...71F} \citeyear{2017ApJ...850...71F}, \citeyear{2024ApJ...961..162F}; SNR RCW~86, \citeauthor{2017JHEAp..15....1S} \citeyear{2017JHEAp..15....1S}, \citeyear{2019ApJ...876...37S}; SN1006, \citeauthor{2022ApJ...933..157S} \citeyear{2022ApJ...933..157S}), correcting for H{\sc i} self-absorption becomes essential for accurately estimating the total interstellar protons.

\bibliography{references}{}
\bibliographystyle{aasjournal}
\end{document}